\documentclass[aps,prd,twocolumn,superscriptaddress,showpacs]{revtex4}
\usepackage{graphicx}
\usepackage{dcolumn}
\usepackage{bm}
\usepackage{natbib}

\newcommand{\apjl}{{Astrophys.~J.~Lett.}}
\newcommand{\apjs}{{Astrophys.~J.~Supp.}}
\newcommand{\aj}{{Astron.~J.}}
\newcommand{\mnras}{{Mon.~Not.~R.~Astron.~Soc.}}

\newcommand{\beq}{\begin{equation}}
\newcommand{\eeq}{\end{equation}}
\newcommand{\beqa}{\begin{eqnarray}}
\newcommand{\eeqa}{\end{eqnarray}}
\newcommand{\bi}{\bf}

\begin{document}

\title{Can superhorizon cosmological perturbations explain the 
acceleration of the universe?}

\author{Christopher M. Hirata}
\email{chirata@princeton.edu}
\affiliation{Department of Physics, Jadwin Hall, Princeton University, 
  Princeton, New Jersey 08544, USA}

\author{Uro\v s Seljak}
\affiliation{Department of Physics, Jadwin Hall, Princeton University, 
  Princeton, New Jersey 08544, USA}
\affiliation{International Center for Theoretical Physics,
  Strada Costiera 11, 34014 Trieste, Italy}

\date{March 27, 2005}

\begin{abstract}
We investigate the recent suggestions by Barausse et al. 
(astro-ph/0501152) and Kolb et al. (hep-th/0503117) that the acceleration 
of the universe could be explained by large superhorizon fluctuations 
generated by inflation. We show that no acceleration can be produced by 
this mechanism.  We begin by showing how the application of Raychaudhuri 
equation to inhomogeneous cosmologies results in several ``no go'' 
theorems for accelerated expansion. Next we derive an exact solution for a 
specific case of initial perturbations, for which application of the Kolb 
et al. expressions leads to an acceleration, while the exact solution 
reveals that no acceleration is present. We show that the discrepancy can 
be traced to higher order terms that were dropped in the Kolb et al. 
analysis. We proceed with the analysis of initial value formulation of 
general relativity to argue that causality severely limits what observable 
effects can be derived from superhorizon perturbations. By constructing a 
Riemann normal coordinate system on initial slice we show that no infrared 
divergence terms arise in this coordinate system. Thus any divergences 
found previously can be eliminated by a local rescaling of coordinates and 
are unobservable. We perform an explicit analysis of the variance of the 
deceleration parameter for the case of single field inflation using usual 
coordinates and show that the infrared divergent terms found by Barausse 
et al. and Kolb et al. cancel against several additional terms not 
considered in their analysis.  Finally, we argue that introducing 
isocurvature perturbations does not alter our conclusion that the 
accelerating expansion of the universe cannot be explained by superhorizon 
modes.
\end{abstract}


\maketitle

\section{Introduction}

There are now several lines of evidence pointing toward an acclerating 
expansion of the universe.  These include the luminosity distance-redshift 
relation measured from Type Ia supernovae (SN Ia) 
\cite{1998AJ....116.1009R, 1999ApJ...517..565P,
2003ApJ...594....1T, 2003ApJ...598..102K, 
2004ApJ...602..571B, 2004ApJ...607..665R}; the 
combination of the angular diameter distance to the surface of last 
scattering and the physical matter density $\Omega_mH_0^2$ measured from 
the cosmic microwave background (CMB) with the low values of $\Omega_mH_0$ 
favored by large-scale structure data \cite{2003ApJS..148....1B, 
2002MNRAS.335..432V, 2004ApJ...606..702T, 2004PhRvD..69j3501T, 
2005astro.ph..1171E}; and, most 
recently, the integrated Sachs-Wolfe effect 
\cite{2003ApJ...597L..89F,2004MNRAS.350L..37F, 2003astro.ph..7335S,
2004PhRvD..69h3524A,
2004Natur.427...45B, 2004ApJ...608...10N,2004astro.ph..4348B,
2004astro.ph.10360P}.  
It is well-known that such an 
accelerating expansion is impossible if one makes the following three 
assumptions:
\newcounter{L1}
\begin{list}{\arabic{L1}}{\usecounter{L1}}
\item\label{it:sec}) the strong energy condition (SEC) holds, i.e. the 
density and isotropic part of the pressure seen by all observers on 
timelike trajectories satisfy $\rho+3p\ge 0$;
\item\label{it:gr}) the universe is described by general relativity (GR); 
and
\item\label{it:frw}) the universe is homogeneous and isotropic, in 
particular the 
Friedmann-Robertson-Walker (FRW) metric is applicable.
\end{list}
Any explanation for the acceleration of the universe must drop at least 
one of these three assumptions.  Usually either assumption 
\#\ref{it:sec} or \#\ref{it:gr} is dropped.  In such a case, we use the 
term ``dark energy'' to describe any SEC-violating matter field, and 
``modified gravity'' to denote any departure from GR.  These explanations 
for the acceleration could be considered unsatisfying since there is 
presently no other 
experimental motivation for modifications to GR, and the matter fields 
normally considered in 
cosmology, including baryonic matter, photons, neutrinos, and cold dark 
matter (CDM) all obey the SEC.  In either case, new physics must be 
invoked.  In contrast, it is observed that 
assumption \#\ref{it:frw} is not exactly valid in the real universe.  
Therefore several recent papers \cite{2005astro.ph..1152B, 
2005hep.th....3117K} have asked whether in fact the ``backreaction'' from 
these perturbations to the universe can explain the accleration, without 
dropping the SEC or GR.

The purpose of this paper is to examine the recent suggestions by Barausse 
et~al. \cite{2005astro.ph..1152B} (hereafter BMR) and Kolb et~al. 
\cite{2005hep.th....3117K} (hereafter KMNR) that perturbations on scales 
larger than the Hubble length can explain the acceleration.  In particular, 
these papers suggest that the time evolution of these large-scale 
perturbations produce a large variance of the deceleration parameter $q$.  
Since potential perturbations at the horizon scale are of order $\sim 
10^{-5}$, one would expect the fluctuations in $q$ to be of this order, 
however KMNR argues that corrections due to very large-scale modes 
(hundreds of $e$-folds outside the horizon) can cause the standard 
deviation of $q$ to be $\gg 10^{-5}$.  In particular, for spectral index 
$n_s\le 1$ they claim that the corrections from very large-scale modes 
contain an infrared divergence. If the variance is very large, this could 
mimick dark energy and cause an apparent acceleration.  Indeed, the 
existence of perturbations on scales well beyond the horizon is likely in 
the context of inflation \cite{1981PhRvD..23..347G, 1982PhLB..108..389L, 
1982PhRvL..48.1220A}, since the inflaton field may have been in the ``slow 
roll'' regime for many more $e$-folds than the minimum of $\sim 60$ 
necessary to solve the horizon problem.  This paper focuses on the 
possibility of accelerated expansion from super-horizon perturbations; we 
do not consider mechanisms based on sub-horizon physics here.

We argue that in fact no acceleration is produced.  Our arguments can be 
summarized as follows:
\begin{list}{$\bullet$}{}
\item The result that GR+SEC forbids accelerating expansion is more 
general than the FRW cosmology.  The application of the Raychaudhuri 
equation to inhomogeneous cosmologies immediately yields several ``no go'' 
theorems for accelerating expansion with zero vorticity, depending on how 
acceleration is defined.  One of these theorems rules out the KMNR 
mechanism.  We treat the definition of the deceleration parameter $q$ and 
the conditions under which $q<0$ is possible in Sec.~\ref{sec:decel}.
\item In Sec.~\ref{sec:accel}, we examine the physical reason why
KMNR found that their deceleration 
parameter becomes negative, in contradiction to our conclusion in 
Sec.~\ref{ss:conditions}.  In particular, we examine one configuration of 
initial perturbations that can be solved exactly;
we show that in this case, the discrepancy arises because of 
higher-order terms in the perturbative expansion of $\Psi$ (their Eq.~7) 
that were dropped in their analysis.  Once the full result for $\Psi$ is 
used, the deceleration parameter $q$ never becomes negative.  The results 
of Sec.~\ref{sec:accel} are less general than those of 
Sec.~\ref{sec:decel}, since they only apply to a specific configuration.  
Nevertheless they provide important insight into the calculation of KMNR.
\item Causality severely limits what observable effects can be 
derived from superhorizon perturbations.  If one constructs an initial 
3-dimensional hypersurface $\Sigma$ at the end of inflation, then any 
observable quantities can in principle be determined purely from the 
values of the initial perturbations on $\Sigma\cap J^-({\cal O})$, where 
$J^-({\cal O})$ is the causal past of the observer.  There may be 
perturbations on very large scales, e.g. due to the early stages of 
inflation, but one can only observe these to the extent that they affect 
the initial conditions on $\Sigma\cap J^-({\cal O})$.  We show in the 
first half of Sec.~\ref{sec:causal} that the observable effect of 
superhorizon perturbations found by BMR -- including the infrared 
divergence -- is in 
fact due to a local rescaling of the coordinates caused by the particular 
choice of gauge for the superhorizon modes.
\item In Sec.~\ref{ss:ircancel}, we calculate the contribution of very 
long-wavelength modes to the deceleration parameter in single-field 
inflation models.  We find that this contribution vanishes when all 
effects, including non-Gaussianity and loop corrections to the scalar 
power spectrum, are taken into account.  This means that the 
long-wavelength modes in single-field 
inflationary models cannot produce a large variance of the deceleration 
parameter.
\end{list}
Sec.~\ref{sec:decel} shows that regardless of the 
potential perturbations, ``acceleration'' as defined by KMNR cannot be 
obtained without sacrificing the SEC or GR or invoking vorticity (which
is not present in KMNR and is not predicted by inflation), while 
Sec.~\ref{sec:causal} 
argues that there is no infrared divergence and that the fluctuations in 
$q$ at the horizon scale are indeed of order $\sim 10^{-5}$.  
Secs.~\ref{sec:accel} and \ref{ss:ircancel} deal with the more specific 
issues of why KMNR found acceleration and infrared divergence, 
respectively, and address technical points about their calculation.  Note 
that our arguments involving the ``no go'' theorems are logically 
independent from the arguments involving causality.  Either one rules out 
the KMNR mechanism as the explanation for the accelerating expansion of 
the universe.

The results of Sec.~\ref{ss:ircancel} are specific to single-field 
inflation, and do not apply to other models.  For example, in multi-field 
inflation large isocurvature perturbations are possible, which can produce 
large variances in the deceleration parameter when one averages over 
different portions of the universe \cite{2004astro.ph.10541K}.  However 
the analysis of Sec.~\ref{sec:decel} still applies to these cases, i.e. 
for several of the definitions of the deceleration parameter there is no 
possibility of obtaining $q<0$ without rejecting the SEC or GR.

The notation used in this paper is as follows: the metric signature will 
be $-+++$, with $h_{ij}$ denoting the spatial metric.  Greek indices will 
range over $\{0,1,2,3\}$ whereas Latin indices range only over the spatial 
directions $\{1,2,3\}$.  In cases where we 
do explicit calculations with the metric, we will use the synchronous 
CDM-comoving gauge, in which $g_{00}=-1$, $g_{0i}=0$, and the CDM 
particles move along curves of constant spatial coordinates $x^k$ (we show 
in Sec.~\ref{sec:decel} that this gauge exists in any scenario, including 
KMNR, where the CDM vorticity vanishes). Symmetrization of indices will be 
defined by $A_{(ab)}=\frac{1}{2}(A_{ab}+A_{ba})$, and similarly for 
antisymmetrization, which is denoted with square brackets $A_{[ab]}$.  In 
the model of KMNR, there are only scalar 
perturbations present initially, and the spatial metric becomes
\beq
h_{ij}(t,x^k)\rightarrow a^2(t) e^{-10\varphi(x^k)/3} \delta_{ij}
\label{eq:vpdef}
\eeq
at early times when the perturbation mode of interest is outside the 
horizon (but after the end of inflation); we use this as the definition of 
$\varphi$.  The overdot will denote derivatives with respect to the 
physical time coordinate $t$; we will not do any calculations with the 
conformal time in this paper.  The analyses of Sec.~\ref{sec:accel} and 
Sec.~\ref{sec:causal} are done on a flat FRW background with a scale factor 
$a(t)\propto t^{2/3}$ normalized to $a(t_0)=1$ today.

\section{Deceleration parameter}
\label{sec:decel}

In order to make a statement about the universe accelerating, it is 
necessary to have a precise definition of the deceleration parameter that 
makes sense in non-FRW cosmologies.  In the recent backreaction papers, 
several definitions have been proposed.  We consider these here.

\subsection{Definitions}

The simplest definition is based on the congruence of timelike geodesics 
formed by the CDM particles with 4-velocity $u^\mu$; one may 
define a local Hubble constant
\beq
H_1 = \frac{1}{3}\nabla_\mu u^\mu
\label{eq:h1}
\eeq
and then a deceleration parameter
\beq
q_1 = -1-H_1^{-2}u^\mu\nabla_\mu H_1.
\label{eq:q1}
\eeq
Here $u^\mu\nabla_\mu$ is the Lagrangian proper time derivative $d/dt_{\rm 
proper}$ associated with a CDM particle.  Also Eq.~(\ref{eq:h1}) implies 
that $H_1$ is related to the expansion tensor $\theta_{\mu\nu}$ via 
$\theta\equiv \theta^\mu_\mu=3H_1$.  Since the CDM particles follow 
geodesics, the expansion tensor is purely spatial in the CDM 
particle's frame, i.e. it is orthogonal to ${\bi u}$: 
$\theta_{\mu\nu}u^\nu=\theta_{\mu\nu}u^\mu=0$ 
(c.f. Eq~9.2.2 of \cite{1984gere.book.....W}).  We 
will need to decompose $\theta_{\mu\nu}$ into the usual irreducible 
tensors 
\beq
\theta_{\mu\nu} = \frac{1}{3}\theta(g_{\mu\nu}+u_\mu u_\nu)
+ \omega_{\mu\nu} + \sigma_{\mu\nu},
\eeq
where $g_{\mu\nu}+u_\mu u_\nu$ is the spatial metric (where ``spatial'' is 
defined by an observer sitting on a CDM particle),
$\omega_{\mu\nu}=\nabla_{[\mu}u_{\nu]}$ is a 
$3\times 3$ purely spatial antisymmetric tensor with 3 independent 
components, and $\sigma_{\mu\nu}$ is a traceless-symmetric spatial tensor 
with 5 independent components.

In the special case where the vorticity 2-form $\omega= 
\frac{1}{2}d{\bi u}$ is zero, we can integrate the exact 1-form ${\bi u}$ 
and write ${\bi u}=-dt$ for some locally defined 0-form $t$.  (In most 
cases of interest, the spacetime is simply connected and $t$ is defined 
globally.)  The congruence of CDM particles is then said to be 
irrotational.  If one labels the CDM particles with coordinates $x^i$, 
then ${\bi u}=-dt$ implies that the covariant components of ${\bi u}$ are 
$u_\mu=(-1,0,0,0)$.  However since the particle spatial coordinates are 
constant, $u^i=0$; normalization $u_\mu u^\mu = -1$ requires $u^0=1$ and 
$u^\mu=(1,0,0,0)$.  The requirement that these components of $u_\mu$ and 
$u^\mu$ be consistent forces $g_{00}=-1$ and $g_{0i}=0$, and the line 
element of the metric then becomes
\beq
ds^2 = -dt^2 + h_{ij}(x^k,t)\,dx^idx^j.
\label{eq:syncom}
\eeq
A straightforward computation shows that
\beq
H_1 = \frac{1}{3}\Gamma^\mu_{\mu\nu}u^\nu
= \frac{1}{6}h^{ij}\partial_th_{ij}
= \frac{1}{6}\partial_t\ln\det{\bf h},
\eeq
and this is recognized as the definition of the variable ``$H(t,x^i)$'' 
used by Ref.~\cite{2004astro.ph.10541K}.  It is also equivalent to the 
operational definition used by KMNR since our $\det{\bf h}$ is equivalent 
to their $\bar a^6e^{-6\Psi_s}$
and they ignore the small-scale perturbation modes $\Psi_s$.  In the 
latter case where a deceleration parameter is used, it is defined as
\beq
-\frac{{\bar a}\partial_t^2{\bar a}}{(\partial_t{\bar a})^2} = 
-1-\frac{\partial_t{H}_1}{H_1^2} = q_1,
\eeq
i.e. the deceleration parameter $\bar{q}$ used by
KMNR is equivalent to our $q_1$.

The Hubble and deceleration parameters $H_1$ and $q_1$ as defined above 
are functions of both the time coordinate $t$ and the spatial coordinates 
$x^i$.  It is sometimes proposed to measure spatially averaged values of 
the Hubble constant,
\beq
H_2(t) = \frac{\int_{\cal V} H_1 \sqrt{h}\,d^3x}{\int_{\cal V} 
\sqrt{h}\,d^3x} = \frac{1}{3}\partial_t\ln\int_{\cal V} \sqrt{h}\,d^3x
\eeq
and
\beq
q_2(t) = \frac{\int_{\cal V} q_1 \sqrt{h}\,d^3x}{\int_{\cal V}
\sqrt{h}\,d^3x}.
\eeq
Note that, unlike the case of $q_1$, we have $q_2\neq -1-\dot{H_2}/H_2^2$.

The deceleration parameters $q_1$ and $q_2$ are convenient because they are 
easy to compute in many cosmological models, avoiding the complicated 
process of tracing null geodesics and estimating luminosity distances used 
by BMR.  The latter, of course, is their principal shortcoming: $q_1$ and 
$q_2$ do not correspond to the observables in any of today's major 
cosmological probes.  The famous Hubble plot of $z$ versus $d_L$ allows 
one to measure $H_1$, however extragalactic astronomy is only $\sim 10^2$ 
years old and this is a woefully inadequate baseline with which to 
directly measure the time derivative $\dot{H}_1$ and hence the 
deceleration parameter $q_1$.  We also cannot cover the entire universe 
with astronomers so as to measure spatially averaged quantities such as 
$H_2$ or $q_2$.

Since SN Ia luminosity distances are one of the major pieces of evidence 
for the accelerating expansion of the universe, one way to make contact 
with observations is to calculate the luminosity distance-redshift 
relation.  This was first done in the context of backreaction by BMR, who 
expanded the luminosity distance $d_L$ as a Taylor series in redshift,
\beq
d_L = H_{3'}^{-1}z + \frac{1-q_{3'}}{2H_{3'}}z^2 + O(z^3).
\label{eq:dlz}
\eeq
As defined here, the Hubble and deceleration parameters $H_{3'}$ and 
$q_{3'}$ depend not only on the observer's spacetime coordinates $t,x^i$, 
but also on the direction of observation (RA and Dec), because in a 
perturbed universe there is no reason why the $d_L$-$z$ relation should be 
the same in every direction.  BMR then went on to average these over solid 
angle to obtain
\beq
H_3 = \langle H_{3'}\rangle_{4\pi} {\rm ~~and~~}
q_3 = \langle q_{3'}\rangle_{4\pi}.
\eeq
A related procedure is to write Eq.~(\ref{eq:dlz}) as a function
$z(d_L)$,
\beq
z = H_{3'}d_L - \frac{1}{2}H_{3'}^2(1-q_{3'})d_L^2 + O(d_L^3).
\label{eq:q3p}
\eeq
This motivates the definition
\beq
\langle z\rangle_{4\pi} = H_4d_L - \frac{H_4^2(1-q_4)}{2}d_L^2 + O(d_L^3),
\label{eq:q4}
\eeq
where we take the angular average over the redshift at fixed 
$d_L$ rather than averaging the Taylor coefficients.
This procedure has the advantage of making sense in slightly more general 
conditions.  In particular, if $\theta_{ij}$ has a
direction $n^i$, not necessarily an eigenvalue, where
$\theta_{ij}n^in^j=0$, then $H_{3'}=0$ in that direction and $q_{3'}$ in 
Eq.~(\ref{eq:dlz}) is undefined.  Then $q_3$ is not defined either.  In 
contrast, Eq.~(\ref{eq:q4}) breaks down only under the more restricted 
condition that $H_4=0$, i.e. that the angular average of $H_{3'}$ 
vanishes.  As we will see in the next section, it is also easier to prove 
theorems about $q_4$ because one can avoid angular averages 
with quantities such as $H_{3'}$ in the denominator.

One can relate $H_4$ and $q_4$ to $H_3$ and $q_3$ by considering the 
angular average of the coefficients of $d_L$ and $d_L^2$ in 
Eq.~(\ref{eq:q3p}),
\beqa
H_4 &=& \langle H_{3'}\rangle_{4\pi} = H_3 {\rm ~~and}
\nonumber \\
H_4^2(1-q_4) &=& \langle H_{3'}^2(1-q_{3'})\rangle_{4\pi}.
\label{eq:14}
\eeqa
Thus we always have $H_3=H_4$.  If $H_{3'}$ is isotropic, then it cancels 
out in the second line implying $q_3=q_4$, but $q_3$ and $q_4$ need not be 
identical in general.

\subsection{Conditions for acceleration}
\label{ss:conditions}

Having written down several definitions of the deceleration parameter, one 
can ask: under what conditions can the universe be accelerating, i.e. 
under what conditions is $q<0$ possible?

The condition for acceleration is easiest to derive in the case of $q_1$.  
We begin with the 
Raychaudhuri equation, 
\beq
{d \theta \over d t}=-{\theta^2 \over 3}-\sigma_{\mu \nu}\sigma^{\mu \nu}
+\omega_{\mu \nu}\omega^{\mu \nu}-R_{\mu\nu}u^\mu u^\nu,
\eeq
where $R_{\mu\nu}$ is the Ricci tensor.  
Using this
Eq.~(\ref{eq:q1}) can be 
re-written as
\beq
H_1^2q_1 = 
\frac{1}{3}(\sigma^\mu_\nu\sigma^\nu_\mu-\omega^\mu_\nu\omega^\nu_\mu) + 
\frac{1}{3}R_{\mu\nu}u^\mu u^\nu.
\eeq
The Einstein 
equation allows us to re-write the last term in terms of the matter 
content in the CDM particles' rest frame,
\beq
H_1^2q_1 = 
\frac{1}{3}(\sigma^\mu_\nu\sigma^\nu_\mu-\omega^\mu_\nu\omega^\nu_\mu) + 
\frac{4\pi G}{3}(\rho + 3p),
\label{eq:hq1}
\eeq
where $\rho=T^{00}$ is the total density and $p=\frac{1}{3}T^i_i$ is the 
isotropic pressure.  Since $\sigma$ is purely spatial in the local Lorentz 
frame of a CDM particle, its square is semipositive definite, 
$\sigma^\mu_\nu\sigma^\nu_\mu\ge 0$.
Thus we immediately see that if the strong energy condition (SEC) is 
satisfied (so that $\rho+3p\ge 0$) and there is no vorticity ($\omega=0$), 
it follows that $q_1\ge 0$.

The vorticity $\omega$ plays a key role in all of the ``no go'' theorems 
regarding accelerated expansion, as well as in the existence of the 
synchronous CDM-comoving gauge (Eq.~\ref{eq:syncom}) and the spatially 
averaged deceleration $q_2$, so it is worth exploring when it can be 
nonzero.  It can be shown (see e.g. Eq.~9.2.14 of 
\cite{1984gere.book.....W}) that if $\omega=0$ at one time then it remains 
zero at all future times (at least until orbit crossings appear).  In 
particular, the existence of the synchronous CDM-comoving gauge implies 
${\bi u}=-dt$ and $\omega=d{\bi u}=-d^2t=0$, so if the synchronous 
CDM-comoving gauge exists at early times then $\omega=0$.  KMNR explicitly 
use the synchronous CDM-comoving gauge in their calculations, so their 
model has zero vorticity.  This means that the ``no go'' theorems that 
$q_1\ge 0$ applies to KMNR; since they use $q_1$ as their 
definition for the deceleration parameter, the KMNR mechanism cannot 
produce acceleration.  A corollary is that $q_1\rightarrow -1$ in KMNR 
must be an artifact of the perturbative expansion.  We will return to this 
point in Sec.~\ref{sec:accel}.

Since $q_2$ is a spatial average of $q_1$, it follows trivially that in 
the absence of vorticity or SEC violation, $q_2\ge 0$.

The deceleration parameters $q_3$ and $q_4$ in terms of luminosity 
distances can be related to observations.  The simplest way to do this is 
to consider the photon trajectories that arrive at the observer ${\cal 
O}$, and find the luminosity distance $d_L(v)$ and redshift $z(v)$ as a 
function of affine parameter $v$.  The photon is assumed to arrive at the 
observer at affine parameter $v=0$ with unit energy, i.e. its 4-momentum 
${\bi k}$ satisfies $k_\mu u^\mu=-1$ at $v=0$.  We will assume that the 
voriticity of the CDM particles vanishes so that we can use the 
synchronous CDM-comoving gauge (Eq.~\ref{eq:syncom}).  In this gauge, the 
``0'' direction coincides with the CDM particle 4-velocity, so the purely 
spatial nature of the expansion tensor $\theta_{\mu\nu}$ can be used: 
$\theta_{00}=\theta_{0i}=\theta_{i0}=0$, and we will henceforth simply 
write $\theta_{ij}$.  We also define the photon's 3-velocity at the 
observer $n_i = k_i(v=0)$, and note that this is a unit vector since the 
photon is massless.  Note that $v<0$ on the observer's past light cone.

The redshift associated with a particular point on the photon's trajectory 
is simply $1+z(v)=k^0(v)$, since $k^0(v)$ is the 
photon's energy as measured by the CDM particle and the observer sees 
energy $1$.  We can expand $k^0(v)$ to second order using the 
geodesic equation,
\beq
\frac{dx^\alpha}{dv} = k^\alpha {\rm ~and~}
\frac{dk^\alpha}{dv} = -\Gamma^\alpha_{\mu\nu}k^\mu k^\nu.
\eeq
The derivatives of $k^0$ are then
\beq
\frac{dk^0}{dv} = -\Gamma^0_{ij}k^ik^j = -\frac{1}{2}\dot{h}_{ij}k^i 
k^j
\eeq
since $\Gamma^0_{00}=\Gamma^0_{0i}=0$ in the synchronous gauge, and
\beq
\frac{d^2k^0}{dv^2} = 
-\frac{1}{2}k^\alpha (\partial_\alpha\dot{h}_{ij})k^ik^j
+ \dot{h}_{ij}k^i\Gamma^j_{\mu\nu}k^\mu k^\nu.
\eeq
Plugging in the Christoffel symbols yields
\beqa
k^0 &=& 1 + \frac{dk^0}{dv}v + 
\frac{d^2k^0}{dv^2}\frac{v^2}{2}
\nonumber \\
&=& 1 - \frac{1}{2}\dot{h}_{ij}n^in^j v + \Biggl[
-\frac{n^in^j}{2}(\ddot{h}_{ij}+n^k\dot{h}_{ij,k})
\nonumber \\ &&
+ \frac{\dot{h}_{ij}n^j}{2}(
2h^{ik}\dot{h}_{kl}n^l - h^{il}h_{mk,l}n^kn^m
\nonumber \\ &&
+ 2h^{il}h_{ml,k}n^kn^m
)\Biggr]\frac{v^2}{2}.
\eeqa
Introducing the extrinsic curvature $K_{ij}=-\frac{1}{2}\dot{h}_{ij}$, 
and its 3-dimensional covariant derivative $K_{ij|k}$,
\beq
K_{(ij|k)} = K_{(ij,k)} + K{^l}_{(i}h_{jk),l} - 2h_{l(j,k}K{^l}_{i)},
\eeq
yields the redshift
\beq
z = K_{ij}n^in^j v + 
\left(\dot{K}_{ij} + K_{ij|k}n^k + 4K^k_iK_{kj}
\right)n^in^j\frac{v^2}{2}.
\label{eq:z4}
\eeq
Since ${\bi u}$ is the unit normal to the hypersurfaces of constant $t$, 
it follows that $K_{ij}=-\theta_{ij}$ (c.f. Eq.~9.3.19 of 
\cite{1984gere.book.....W}).

The luminosity distance to a spherical source whose surface is at affine 
parameter $v_s$ is given by Eq.~(2.14) of 
Ref.~\cite{1987MNRAS.228..653S}
\beq
d_L(v_s) = R\frac{A(v_s)}{A(0)}[1+z(v_s)],
\eeq
where $A$ is the photon amplitude (normalized by $T^{\mu\nu}=A^2k^\mu 
k^\nu$) and $R$ is the radius of the source; the amplitude $A$ is given 
by
\beq
A\propto\exp\left(-\frac{1}{2}\int\hat\theta\,dv\right),
\label{eq:a}
\eeq
where $\hat\theta=\nabla_\mu k^\mu$ is the photon expansion.  The photon 
expansion near the source can be expanded as
\beq
\hat\theta = \frac{2}{v-(v_s+\Delta v_s)} + O[v-(v_s+\Delta v_s)],
\label{eq:thetahat}
\eeq
where $-\Delta v_s$ is the difference affine parameter from the source's 
center to its surface at radius $R$; this can be shown by the same 
argument leading to BMR's Eq.~(A27).  This affine parameter can be found 
by noting that $k^\mu = dx^\mu/dv$.  Here the photon's energy at 
the source is $k^0=1+z$ and the time it takes to travel from the center to 
the surface is $R$, so
\beq
\Delta v_s = -\frac{k^0}{\Delta t} = -\frac{1+z}{R}.
\eeq
Also, integration of Eq.~(\ref{eq:a}) using Eq.~(\ref{eq:thetahat}) gives
\beq
\frac{A(v_s)}{A(0)} = \frac{-\Delta v_s}{-v_s-\Delta v_s}
= \frac{1+z}{-v_sR},
\eeq
where the last equality uses $|\Delta v_s|\ll |v_s|$.
This yields
\beq
d_L(v_s) = -v_s[1+z(v_s)]^2+O(v_s^3).
\eeq
Comparing to Eq.~(\ref{eq:z4}) gives the result, to second order, that
\beq
z = -\frac{K_{ij}n^in^j d_L}{(1+z)^2} +
\left(\dot{K}_{ij} + K_{ij|k}n^k + 4K^k_iK_{kj}
\right)n^in^j\frac{d_L^2}{2},
\eeq
or
\beqa
z &=& -K_{ij}n^in^jd_L - 2(K_{ij}n^in^j)^2d_L^2
\nonumber \\ && +
\left(\dot{K}_{ij} + K_{ij|k}n^k + 4K^k_iK_{kj}
\right)n^in^j\frac{d_L^2}{2}.
\label{eq:zz4}
\eeqa

We can then calculate the deceleration parameter $q_4$ by comparing 
Eq.~(\ref{eq:zz4}) to Eq.~(\ref{eq:q4}).  The $K_{ij|k}$ term drops out in 
the angular averaging, and we have $\langle n^in_i\rangle_{4\pi} = 
\frac{1}{3}$, so we get
\beq
H_4 = -\frac{1}{3}K
\eeq
where $K=K^i_i$.  Note that this implies $H_4=H_1$.  The deceleration 
parameter is
\beqa
q_4 &=& 1 + H_4^{-2}\Biggl[-4K_{ij}K_{kl}\frac{h^{(ij}h^{kl)}}{5}
+ \frac{1}{3}\dot{K}_{ij}h^{ij}
\nonumber \\ &&
+ \frac{4}{3}K^i_jK^j_i\Biggr].
\eeqa
This can be simplified by separating $K$ into its isotropic and 
anisotropic pieces, $K^i_j = -H_4\delta^i_j - \sigma^i_j$.  Then we find
\beq
q_4 = 1 + \frac{\dot{K}_{ij}h^{ij}}{3H_4^{2}}
+ \frac{4}{5}H_4^{-2}\sigma^i_j\sigma^j_i.
\label{eq:q4a}
\eeq
A further simplification comes from the formula for $\dot{K}_{ij}$ 
(Eq.~21.82 of Ref.~\cite{1973grav.book.....M}),
\beq
R{^0}_{i0j} = -(\dot{K}_{ij} + K_{ik}K^k_j);
\eeq
the antisymmetry property of the Riemann tensor coupled with the Einstein 
equation allows us to write
\beq
h^{ij}R{^0}_{i0j} = R^0_0 = 8\pi G\left(T^0_0 - \frac{1}{2}T^\alpha_\alpha 
\right) = -4\pi G(\rho+3p),
\eeq
hence
\beq
\dot{K}_{ij}h^{ij} = 4\pi G(\rho+3p) - 3H_4^2 - \sigma^i_j\sigma^j_i.
\eeq
Substitution into Eq.~(\ref{eq:q4a}) yields
\beq
q_4 = \frac{4\pi G(\rho+3p)}{3H_4^2} + 
\frac{7\sigma^i_j\sigma^j_i}{15H_4^2}.
\label{eq:hq2}
\eeq
Once again, we see that if the SEC holds then $q_4\ge 0$.  Thus by this 
definition too a universe with only matter cannot accelerate.

The deceleration parameter $q_3$ is only well-defined for a 
positive-definite spatial expansion tensor $\theta_{ij}$.  It is also more 
difficult to work with than $q_4$: from Eq.~(\ref{eq:q3p}), we find
\beq
q_{3'} = 1+2\frac{(\partial^2 z/\partial d_L^2)|_{d_L=0}}
{(\partial z/\partial d_L|_{d_L=0})^2}.
\eeq
The presence of variables in the denominator of this equation makes it 
very difficult to prove theorems about the angular
average $q_3=\langle q_{3'}\rangle_{4\pi}$.  We do know from
Eq.~(\ref{eq:14}) that $H_4 = H_3$.
Since the mean of the square is always at least the square of the mean, 
this implies
\beq
H_4^2 \le \langle H_{3'}^2\rangle_{4\pi}.
\label{eq:int1}
\eeq
Combining Eq.~(\ref{eq:int1}) with the second line of Eq.~(\ref{eq:14}) 
yields
\beq
\langle H_{3'}^2q_{3'} \rangle_{4\pi} =
\langle H_{3'}^2\rangle_{4\pi} - H_4^2 + H_4^2q_4 \ge H_4^2q_4.
\eeq
If the SEC holds then $q_4\ge 0$ and it follows that $\langle 
H_{3'}^2q_{3'} \rangle_{4\pi} \ge 0$.  This inequality does allow for the 
possibility that $q_3 = \langle q_{3'}\rangle_{4\pi}< 0$.  This can 
happen if there is an anisotropic expansion tensor (i.e. $\sigma\neq 0$) 
so that $\langle
H_{3'}^2q_{3'} \rangle_{4\pi} \neq H_3^2\langle q_{3'}\rangle_{4\pi}$ 
and the universe is ``accelerating'' with $q_{3'}<0$ in some directions 
$n^i$.  Nevertheless even for an anisotropic expansion, the inequality 
$\langle H_{3'}^2q_{3'} \rangle_{4\pi} \ge 0$ implies that if there are 
accelerating directions ($q_{3'}<0$) then there must be decelerating 
directions ($q_{3'}>0$) as well.

\section{Acceleration of underdense regions?}
\label{sec:accel}

KMNR concluded that an underdense region of a post-inflationary universe 
filled with only dark matter can appear to accelerate (according to $q_1$) 
due to the influence of super-horizon perturbations.  However, with CDM 
only the SEC should hold, and after inflation there should be no 
vorticity.  In Sec.~\ref{ss:conditions}, we showed that the SEC and zero 
vorticity imply $q_1\ge 0$.  These conclusions are obviously inconsistent.  
The purpose of this section is to show, in a particular case that can be 
solved exactly, that the reason for the inconsistency is that KMNR dropped 
higher-order perturbative terms in their gravitational potential 
$\Psi_\ell$.  These terms, when incorporated into their Eq.~(10), result 
in a deceleration $q_1$ that never becomes negative.

The particular case that we shall consider here is the potential
\beq
\varphi = \frac{3}{5}\ln \left( 1 - \frac{1}{4}C|x|^2 \right),
\label{eq:open}
\eeq
where $|x|^2 = (x^1)^2+(x^2)^2+(x^3)^2$ and $C$ is a constant.  In order 
for the conclusions derived below to be valid, it is sufficient by 
causality for $\varphi$ to take on the above value inside our horizon.  
Also, by Birkhoff's theorem, if the potential is spherically symmetric and 
takes on the form of Eq.~(\ref{eq:open}) for $|x|$ less than some radius 
$R$, the future evolution at $|x|<R$ is as computed below.

Since for $C>0$ we have $\nabla^2\varphi<0$, the potential 
Eq.~(\ref{eq:open}) can be used as a model for an underdense region. Not 
all underdense regions look like Eq.~(\ref{eq:open}), however this 
potential can be solved and it is conceptually useful for understanding 
why the higher-derivative terms dropped in KMNR are in fact important.  
The spherically symmetric underdensity and overdensity have been studied 
extensively as models of voids in large-scale structure 
\cite{2004MNRAS.350..517S} and the formation and growth of collapsed 
objects \cite{1972ApJ...176....1G} respectively, however these studies 
were not aimed at understanding backreaction.  Our aim here is to 
understand the implications of the spherical underdensity for the KMNR 
mechanism.

Substitution of the potential in Eq.~(\ref{eq:open}) into 
Eq.~(\ref{eq:vpdef}) yields initial conditions for the spatial metric,
\beq
h_{ij}(t,x^k) = \frac{a^2(t)}{\left(1 - \frac{1}{4}C|x|^2\right)^2}
[\delta_{ij}+O(t^{2/3})].
\label{eq:co1}
\eeq
This is simply the metric for a 3-dimensional open universe with 
spatial curvature $-C$ and a coordinate system given by the 
stereographic projection, analytically continued to the case of negative 
curvature.  The future evolution of this universe is equivalent to an 
open FRW universe and thus can thus immediately be written down:
\beq
ds^2 = -dt^2 + \frac{a^2_{\rm open}(t)}{\left(1 - 
\frac{1}{4}C|x|^2\right)^2}dx^idx^i,
\label{eq:co2}
\eeq
where $a_{\rm open}(t)$ satisfies the Friedmann equation for an open 
universe, which has the parametric solution
\beq
\left\{\begin{array}{rl}
a_{\rm open} & \!\! = \Xi\sqrt{C}\;(\cosh\eta-1) \\
t & \!\! = \Xi(\sinh\eta-\eta) \end{array} \right.,
\label{eq:eta}
\eeq
where $\Xi$ is a constant and $\eta$ is the parameter.  Comparison of 
Eqs.~(\ref{eq:co1}) and (\ref{eq:co2}) shows that at early times, $a_{\rm 
open}(t)/a(t)\rightarrow 1$.  Since the evolution of $a(t)$ is simply that 
of the Einstein-de Sitter universe,
\beq
a(t) = \left(\frac{3}{2}H_0t \right)^{2/3},
\eeq
we can use this as the early-time limit of $a_{\rm open}(t)$ to determine 
$\Xi$,
\beq
\Xi = \frac{1}{2}H_0^2 C^{-3/2}.
\eeq

KMNR defined the metric perturbation variable $\Psi$ according to
\beq
ds^2 = -dt^2 + a^2(t) e^{-2\Psi(x^i,t)}dx^idx^i.
\label{eq:kolb-ds2}
\eeq
Comparison of Eq.~(\ref{eq:co2}) with Eq.~(\ref{eq:kolb-ds2}) then allows 
us to determine the full nonperturbative evolution of $\Psi$ for the 
potential $\varphi$ of Eq.~(\ref{eq:open}):
\beq
\Psi(x^i,t) = \ln\left( 1 - \frac{1}{4}C|x|^2 \right)
- \ln\frac{a_{\rm open}(t)}{a(t)}.
\label{eq:po}
\eeq
The first term in this result is independent of time and, as correctly 
argued by KMNR, it does not affect $q_1$.  The
second term can be evaluated by expressing 
$\eta$ in terms of $t$ and hence $a(t)$ in the second line of 
Eq.~(\ref{eq:eta}) and then finding $a_{\rm open}$ using the first line.
The functional form $\eta(t)$ cannot be expressed analytically, but a 
power series can be developed via reversion of series,
\beq
\eta = \left(\frac{6t}{\Xi}\right)^{1/3}
-\frac{1}{60}\left(\frac{6t}{\Xi}\right)
+\frac{1}{1400}\left(\frac{6t}{\Xi}\right)^{5/3} + ...
\eeq
Substitution into the Taylor expansion of $a_{\rm 
open}(t)=\Xi\sqrt{C}(\cosh\eta-1)$ gives
\beqa
a_{\rm open}(t) &=& a(t) + \frac{C}{5H_0^2}a^2(t) - \frac{3C^2}{175H_0^4}
a^3(t)
\nonumber \\ && + O(C^3H_0^{-6}a^4).
\label{eq:aopen-c}
\eeqa
We are interested in the expression for $a_{\rm open}$ as a perturbation 
series in $C$, since $\varphi\rightarrow 0$ when $C\rightarrow 0$.  
Eq.~(\ref{eq:aopen-c}) provides this series, although it turns out to also 
be a power series in $a$.  [The reason for this is that $C$ has units of 
length$^{-2}$, and $H_0$ is the only other quantity containing units of 
length once we have eliminated $\rho_0$ using the Friedmann equation.
Thus $C$ and $H_0$ appear together in the combination $C/H_0^2$
when expanding $a_{\rm open}$.  Furthermore, there is an invariance 
associated with the choice of epoch of the observer: if one places the 
observer at a later time $\xi t$ instead of $t$, then the scale factors 
$a$ and $a_{\rm open}$ are multiplied by $\xi^{-2/3}$ (since $a\propto 
t^{2/3}$), $H_0$ is multiplied by $\xi^{-1}$, and the change in definition 
of comoving length multiplies the ``curvature'' constant $C$ by 
$\xi^{-4/3}$, hence $C/H_0^2$ is multiplied by $\xi^{2/3}$.  The 
invariance with respect to $\xi$ forces the terms in the above expansion 
to take the form $(C/H_0^2)^ja^{j+1}$.  Thus the expansion in $C$ turns 
out to also be an expansion in $a$.]  Substitution into Eq.~(\ref{eq:po}) 
gives
\beqa
\Psi(x^i,t) &=& 
\ln\left( 1 - \frac{1}{4}C|x|^2\right)
- \frac{C}{5H_0^2}a(t)
\nonumber \\ &&
+ \frac{13C^2}{350H_0^4}a^2(t) + O(C^3H_0^{-6}a^3).
\label{eq:psi-open}
\eeqa
Aside from the irrelevant time-independent term, $\Psi(x^i,t)$ depends 
only on the time.  Technically the perturbation series in 
Eq.~(\ref{eq:psi-open}) has a finite radius of convergence of 
$|CH_0^{-2}a|<(\frac{3}{2}\pi)^{2/3}$ (see Appendix~\ref{app:conv}), but 
if one had the full perturbative expansion one could analytically continue 
it to later times.

Our result for the evolution of the gravitational potential should be 
compared to Eq.~(7) of KMNR, which after plugging in 
Eq.~(\ref{eq:open}) yields
\beq
\Psi(x^i,t) = \ln\left( 1 - \frac{1}{4}C|x|^2\right)
- \frac{C}{5H_0^2}a(t)
+ \frac{C^2|x|^2}{60H_0^2}a(t).
\label{eq:k7}
\eeq
The constant in this equation agrees with Eq.~(\ref{eq:psi-open}).  The 
first-order perturbation terms (i.e. $\propto C$) also agree.  Beyond 
first order in $C$, however, two discrepancies appear.  One is the 
spurious term in Eq.~(\ref{eq:k7}) proportional to $C^2$ that has spatial 
dependence.  We will consider $\Psi$ and $q$ the spatial origin $x^i=0$, 
where the spurious term vanishes.  [This is also where the KMNR 
calculation is most likely to be valid, since they neglected terms
containing $(\nabla\varphi)^2$, and $\nabla\varphi=0$ at the origin for 
our specific potential, Eq.~(\ref{eq:open}).]
The other discrepancy is that Eq.~(\ref{eq:k7}) is missing the 
higher-order 
spatially independent terms proportional to $(Ca)^2$, $(Ca)^3$, etc.  The 
$(Ca)^n$ term has $2n$ derivatives in it and presumably are not included 
in Eq.~(7) of KMNR because they dropped ``higher derivative'' terms.  

We finally consider the implications of the difference between 
Eq.~(\ref{eq:po}) and Eq.~(\ref{eq:k7}) for the determination of the 
deceleration parameter $q_1$.  This is given by Eq.~(6) of KMNR,
\beq
q_1 = -1 + \frac{\frac{3}{2} + H^{-2}\ddot{\Psi}}
{\left(1 - H^{-1}\dot{\Psi} \right)^2}.
\label{eq:k6}
\eeq
Here we have dropped the subscript on $\Psi_\ell$ since $\Psi=\Psi_\ell$ 
at the level of approximation in KMNR, and $H$ 
represents the unperturbed Hubble constant.  
According to Eq.~(\ref{eq:k7}), the value of $q_1$ at the spatial origin 
$x^i=0$ is
\beq
q_1({\rm KMNR}) = -1 + \frac{\frac{3}{2} + \frac{1}{10}H_0^{-2}Ca}
{(1+\frac{1}{5}H_0^{-2}Ca)^2}.
\label{eq:q1kmnr}
\eeq
According to the full nonperturbative calculation, and defining $H_{\rm 
open}(t) = \dot a_{\rm open}(t)/a_{\rm open}(t)$, we find
\beq
\begin{array}{rcl}
\dot{\Psi} & = & H(t) - H_{\rm open}(t) {\rm ~~and}\\
\ddot{\Psi} & = & \dot{H}(t) - \dot{H}_{\rm open}(t).
\end{array}
\eeq
Substituting into Eq.~(\ref{eq:k6}) yields
\beq
q_1({\rm true}) = -1 - \frac{\dot{H}_{\rm open}(t)}{H^2_{\rm open}(t)},
\eeq
where $H(t)$ completely drops out if one uses the identity 
$\dot{H}=-\frac{3}{2}H^2$.  We can use Eq.~(\ref{eq:eta}) to find $H_{\rm 
open}(t)$, which gives
\beq
H_{\rm open}(t) = A^{-1}\frac{\sinh\eta}{\cosh\eta-1}
\eeq
and
\beq
q_1({\rm true}) = \frac{1}{1+\cosh\eta}.
\label{eq:q1true}
\eeq

Equations (\ref{eq:q1kmnr}) and (\ref{eq:q1true}) have some features in 
common: they both approach $1/2$ at early times, and as the universe 
expands they decrease, as expected for an underdense region of the 
universe.  However the long-time behavior is different: as 
$t\rightarrow\infty$, $q_1({\rm KMNR})\rightarrow -1$, as noted by KMNR, 
whereas $\eta\rightarrow\infty$ and hence
$q_1({\rm true})\rightarrow 0$. We thus see explicitly that no 
acceleration is possible in this case and that the result found by KMNR is 
a consequence of neglecting higher order terms in their analysis: if the 
full perturbation series had been calculated (and analytically continued 
to late times), KMNR would have found no acceleration.

\section{Superhorizon adiabatic perturbations}
\label{sec:causal}

BMR have recently analyzed the effect of superhorizon adiabatic 
perturbations on the $d_L$-$z$ relation.  In particular they considered 
the value of the isotropically averaged deceleration parameter $q_3$ 
($\langle q_0\rangle_\Omega$ in their notation).  They found a large 
contribution to the variance of $q_3$ coming from the interaction of 
infrared and ultraviolet modes.  In this section we re-derive this result 
by selecting an appropriate new coordinate system.  This method 
illuminates the physical origin of the large ${\rm Var}\,q_3$ found by 
BMR, namely that BMR assumed the ultraviolet modes to be statistically 
homogeneous, isotropic, and Gaussian in their coordinate system with the 
cosmic power spectrum $P_\varphi(k)$.  Because of the infrared 
perturbations, this coordinate system differs from the locally FRW 
coordinate system erected by an observer, and the power spectrum of 
ultraviolet modes seen by a specific observer differs from $P_\varphi(k)$.  
In Sec.~\ref{ss:ircancel}, we consider the $\varphi$ field generated by 
single-field inflation, and show that in this case the BMR infrared 
divergences are canceled by non-Gaussian features and loop corrections to 
the power spectrum.

The KMNR mechanism makes use of this infrared divergence.  Specifically, 
they argue that in some models of inflation, there is a very large 
variance of the deceleration parameter $q_1$ resulting from the 
interaction of infrared and horizon-scale modes.  Within the context of 
single-field inflation, or indeed any mechanism in which small-scale 
perturbations are laid down after the large-scale perturbations have 
exited the horizon and frozen out, there is then no infrared divergence.  
We discuss the implications for KMNR in Sec.~\ref{ss:3117}.

\subsection{Gauge transformation; causality}

The initial state of any system in classical general relativity is 
described by a spatial metric $h_{ij}$ on the initial hypersurface, 
its time derivative, i.e. 
extrinsic curvature $K_{ij}$, and the appropriate variables $X_a$ to 
describe the matter fields.  These quantities are not all independent 
since they must satisfy the energy and momentum constraints.  It is also 
well-known that two distinct sets of initial conditions 
$\{h_{ij},K_{ij},X_a\}$ may actually correspond to the same spacetime 
because they differ by a gauge transformation, which may involve either 
(i) a re-parameterization of the initial hypersurface, or (ii) selection 
of a different spacelike hypersurface as the initial slice.  We will focus 
our analysis on using gauge transformations of type (i) since (ii) is more 
difficult to analyze.

\begin{figure}
\includegraphics[angle=0,width=3.2in]{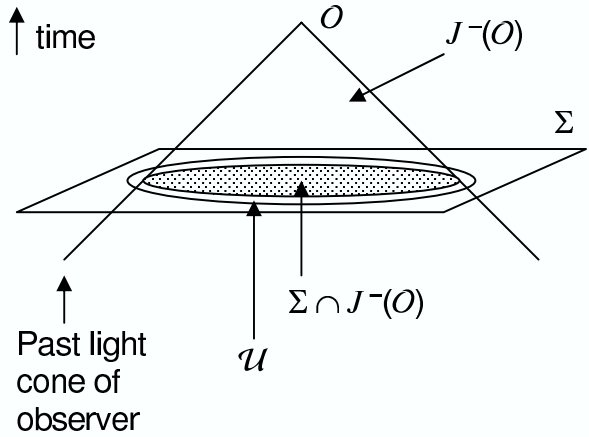}
\caption{\label{fig:lightcone}The regions of spacetime described in the 
text.  Initial data are supplied on the Cauchy hypersurface $\Sigma$.
Causality implies that the observables seen by observer ${\cal O}$ 
can depend only on the initial data within the shaded region $\Sigma\cap 
J^-({\cal O})$.  Within the context of perturbation theory around an FRW 
spacetime, the 
perturbations to the observables can be determined given initial data 
within any open set ${\cal U}$ containing $\Sigma\cap
J^-({\cal O},{\rm FRW})$.  Superhorizon perturbations can only be observed 
to the extent that they affect the initial data inside ${\cal 
U}$.}
\end{figure}

If one considers an initial hypersurface $\Sigma$ after the end of 
inflation, one need only specify the initial data on the portion of 
$\Sigma$ that can be seen by the observer, ${\cal O}$, i.e. one needs 
initial data on $\Sigma\cap J^-({\cal O})$.  In the unperturbed standard 
CDM model with the usual initial hypersurface (a surface of constant 
cosmic time), $\Sigma\cap J^-({\cal O})$ is a closed ball with comoving 
radius less than the horizon size $\chi_h$,
\beq
\chi_h = \int_0^1 \frac{da}{a^2H(a)} = 2H_0^{-1},
\eeq
where we have used $H(a)\propto a^{-3/2}$.  In a non-FRW universe the 
nature of $\Sigma\cap J^-({\cal O})$ is more complicated.  Nevertheless, 
Cauchy stability \cite{1975lsss.book.....H} implies that, for sufficiently 
small perturbations, the 
observables seen by ${\cal O}$ can be uniquely determined from the initial 
data on any open set ${\cal U}$ in $\Sigma$ with ${\cal U}\supseteq 
\Sigma\cap J^-({\cal O},{\rm FRW})$, where $J^-({\cal O},{\rm FRW})$ is 
the causal past of the observer assuming the FRW metric (see 
Fig.~\ref{fig:lightcone}).  In particular 
this means that given such a ${\cal U}$, one can calculate the observables 
seen by ${\cal O}$ to any order in perturbation theory.  It is also easy 
to construct ${\cal U}$: any open ball centered at the origin and 
extending out to coordinate radius $R_{\cal U}>\chi_h$ will do.  We assume
$R_{\cal U}=2\chi_h$ for definiteness.

The initial hypersurface considered by BMR has a spatial metric
\beqa
h_{ij}(t_\Sigma,x^k) &=&
a^2(t_\Sigma)\gamma_{ij}(t_\Sigma,x^k)
\nonumber \\
&=& a^2(t_\Sigma) e^{-10\varphi(x^k)/3} 
[\delta_{ij} + O(t^{2/3})]
\label{eq:gamij}
\eeqa
and extrinsic curvature
\beqa
K_{ij}(t_\Sigma,x^k) &=&
-\frac{1}{2}\dot h_{ij}(t_\Sigma,x^k)
\nonumber \\ &=&
-H(t_\Sigma)h_{ij}(t_\Sigma,x^k) + 
O(K_{ij}t^{2/3}),
\eeqa
where we write $O(K_{ij}t^{2/3})$ to mean that the fractional correction 
is of order $t^{2/3}$.  Instead of $x^i$ coordinates
we want to construct a coordinate system which is 
explicitly locally inertial. 
We will therefore impose a gauge transformation to new coordinates $y^i(x^j)$ such that at early 
times, $h_{ij}$ is very close to $a^2\delta_{ij}$ (without the conformal 
factor).
One realization of such a coordinate system is to construct geodesics 
in all directions from the center and decompose their tangent vectors in 
terms of a chosen basis: the resulting coordinates are the Riemann normal 
coordinates. 

We begin by splitting $\varphi$ in Eq.~(\ref{eq:gamij}) into a large-scale 
and a small-scale piece,
\beq
\varphi(x^i) = \varphi_L(x^i) + \varphi_S(x^i).
\eeq
We will work to first order in $\varphi_S$ and in the wavenumber $k_L$ of 
the large-scale modes.  That is, any terms containing $\varphi_S^2$ or 
$k_L^2\varphi_L\sim \nabla\nabla\varphi_L$ will be dropped.  Note that the 
split into ``long'' and ``short'' wavelengths here differs from the split 
$\Psi=\Psi_\ell+\Psi_s$ in KMNR: we are 
including both, and are doing the separation so that $k_L$ and $\varphi_S$ 
are the expansion parameters, whereas in KMNR 
$\Psi_s$ is dropped.

Next we construct a Riemann normal coordinate system on the initial slice, 
centered at $x^i=0$, with respect to the large-scale metric $\gamma_L$.  
The construction of such a coordinate system begins with the choice of an 
orthonormal basis of three vectors $\{({\bi e}_{\bar a})^i\}_{\bar a=1}^3$ 
tangent to $\Sigma$ at the point $(t_\Sigma,0,0,0)$.  (We use the barred 
indices $\bar{a}$ to denote the components associated with these basis 
vectors and with the ${\bi y}$ coordinate system that will be defined 
based on them.)  Such a basis is not unique; here we choose
\beq
e_{\bar a}{^i} = [\gamma_L^{-1/2}({\bf 0})]_{\bar a}{^i},
\eeq
where one takes the $-1/2$ power of the $3\times 3$ matrix of covariant 
components of $\gamma_{L\,ij}$.  The Riemann normal coordinates are then 
constructed via
\beq
x^i(y^{\bar j}) = \exp_{\bf 0} \left\{ e_{\bar j}{^i} y^{\bar j} \right\},
\eeq
where $\exp$ represents the exponential geodesic mapping.  In our case, 
the argument of the exponential simplifies to the vector
\beq
e_{\bar j}{^i} y^{\bar j} =
[\gamma_L^{-1/2}({\bf 0})]_{\bar j}{^i}y^{\bar j}
= e^{-5\varphi_L({\bf 0})/3}y^i.
\label{eq:init}
\eeq
The geodesic equation of the $\gamma_L$ metric can be found from its 
Christoffel symbols,
\beqa
\frac{d^2x^i}{d\lambda^2} &=& 
-\Gamma^i_{jk}\frac{dx^j}{d\lambda}\frac{dx^k}{d\lambda}
\nonumber \\
&=& \frac{5}{3}e^{10\varphi_L/3} \Biggl( 
  -\frac{\partial\varphi_L}{\partial x^i}
  \left|\frac{dx^j}{d\lambda}\right|^2
\nonumber \\ &&
  + 2\frac{\partial\varphi_L}{\partial x^j}
  \frac{dx^i}{d\lambda}\frac{dx^j}{d\lambda}
  \Biggr).
\eeqa
Solving this equation to second order in $y^i$ subject to the initial conditions ${\bf x}(\lambda=0)={\bf 0}$ and ${\bf 
x}'(\lambda=0)$ given by Eq.~(\ref{eq:init}) gives
\beqa
x^i &=& x^i(\lambda=1)
\nonumber \\ &=&
e^{10\varphi_L({\bf 0})/3}y^i + \frac{5}{6}e^{20\varphi_L({\bf 0})/3} \Biggl[
  -|y|^2\varphi_{L,i}({\bf 0})
\nonumber \\ &&
  + 2y^iy^j\varphi_{L,j}({\bf 0}) \Biggr].
\label{eq:xy}
\eeqa
The metric $\gamma_L$ in Riemann normal coordinates defined with an 
orthonormal basis is $\delta_{\bar i\bar j}$ at $0$ and its first spatial 
derivatives vanish (see e.g. Sec. 11.6 of \cite{1973grav.book.....M}).  
Thus
\beq
\gamma_{L\; \bar i\bar j}(y^{\bar k}) = \delta_{\bar i\bar j} + 
O(k_L^2\varphi_L|y|^2),
\eeq
where $k_L^2\varphi_L$ is the second spatial derivative of $\varphi_L$, 
which is 
suppressed for long-wavelength perturbations.  Note that within the set 
${\cal U}$, $|y|^2$ is at most of order $H_0^{-1}$.
The full metric $\gamma_{\bar i\bar j}=e^{10\varphi_S/3}\gamma_{L\; 
\bar i\bar j}$ is then
\beq
\gamma_{\bar i\bar j}(y^{\bar k}) = e^{10\varphi_S[{\bf x}({\bf 
y})]/3}\delta_{\bar i\bar j} + 
O(k_L^2\varphi_L|y|^2).
\label{eq:metricnew}
\eeq

We may now calculate the deceleration parameter in the new coordinate 
system.  BMR has found the deceleration parameter to be, to first order in 
$\varphi$,
\beq
q_3 = \frac{1}{2} + \frac{5}{9}H_0^{-2}\nabla^2_{\bf 
x}\varphi + O(\varphi^2).
\label{eq:q3-1st}
\eeq
Since Eq.~(\ref{eq:metricnew}) is now simply the Euclidean metric with 
a first-order perturbation, we can apply Eq.~(\ref{eq:q3-1st}) in the 
${\bi y}$ coordinate system to yield
\beq
q_3 = \frac{1}{2} + \frac{5}{9}H_0^{-2}\nabla^2_{\bf 
y}\varphi_S + O(\varphi_S^2,H_0^{-2}k_L^2\varphi_L).
\label{eq:q0}
\eeq
The order of the error is now $H_0^{-2}k_L^2\varphi_L$ since 
Eq.~(\ref{eq:metricnew}) had error $O(k_L^2\varphi_L|y|^2)$ and causality 
requires that we can only observe out to distances $|y|<R_{\cal U}\sim 
O(H_0^{-1})$. Thus
in Riemann normal coordinate system, which is constructed 
locally and does not make reference to anything farther from the observer 
than the horizon, the result for the deceleration parameter is extremely 
simple and large amplitude long 
wavelength modes have no impact on the deceleration parameter.

\subsection{Deceleration parameter in ${\bi x}$ coordinates}

BMR expressed the deceleration parameter $q_3$ in the ${\bi x}$ coordinate 
system rather than ${\bi y}$. We have already shown that the analysis 
in ${\bi y}$ coordinates results in Eq.~\ref{eq:q0} and that 
superhorizon modes cannot affect the deceleration parameter. 
Nevertheless, we wish to compare our result to 
BMR, who did the calculation in the ${\bi x}$ coordinates.
Since the final result for any observable 
cannot depend on the choice of coordinates 
we must obtain the same answer as suggested by 
Eq.~\ref{eq:q0}. We will show
below that this is indeed the case and is a result of a precise cancellation
between several terms, some of which were not considered by BMR or KMNR. 

We can recover all terms in the deceleration parameter up to order 
$k_L\varphi_L$ by using Eq.~(\ref{eq:xy}) to turn $\nabla^2_{\bf 
y}\varphi_S$ into derivatives with respect to ${\bf x}$:
\beq
\nabla^2_{\bf y}\varphi_S = -\frac{5}{3}e^{20\varphi_L/3}\varphi_{L,i}\varphi_{S,i}
  + e^{10\varphi_L/3}\nabla^2_{\bf x}\varphi_S.
\eeq
Plugging into Eq.~(\ref{eq:q0}) yields
\beqa
q_3 &=& \frac{1}{2} + \frac{5}{9}H_0^{-2}\Biggl(
-\frac{5}{3}e^{20\varphi_L/3}\varphi_{L,i}\varphi_{S,i}
\nonumber \\ &&
  + e^{10\varphi_L/3}\nabla^2_{\bf x}\varphi_S
\Biggr)
\nonumber \\ &=&
\frac{1}{2} + \frac{2}{H_0^2}\Biggl(
  \frac{5}{18}\nabla^2_{\bf x}\varphi_S  -\frac{25}{54}\varphi_{L,i}\varphi_{S,i}
\nonumber \\ &&
  + \frac{25}{27}\varphi_L\nabla^2_{\bf x}\varphi_S \Biggr),
\label{eq:q0y}
\eeqa
where in the last line we have worked to order $\varphi_L$.  This should 
be compared to Eq.~(36) of BMR, which reads
\beqa
q_3 &=& \frac{1}{2} + \frac{2}{H_0^2}\Biggl(\frac{5}{18}\nabla_{\bf 
x}^2\varphi
- \frac{25}{54}|\nabla_{\bf x}\varphi|^2 
\nonumber \\ &&
+ \frac{25}{27}\varphi\nabla_{\bf x}^2\varphi
\Biggr) + O(H_0^{-4}).
\eeqa
In particular, the infrared-divergent term $\varphi\nabla_{\bf 
x}^2\varphi$ identified by BMR appears in 
Eq.~(\ref{eq:q0y}).  The $\varphi_{L,i}\varphi_{S,i}$ term is not 
infrared-divergent (unless $n_s\le-1$), nevertheless it is first-order in 
$k_L$ so it appears here; the coefficient is double that in 
BMR because of the factor of 2 in the 
binomial expansion of $(\nabla\varphi)^2$.  The variance of the 
divergent term is
\beqa
{\rm Var}(\varphi\nabla_{\bf x}^2\varphi) \!\! &=& \!\!
\left[\int P_\varphi(k)\, \frac{d^3\bi k}{(2\pi)^3}\right]
\left[\int k^4P_\varphi(k)\, \frac{d^3\bi k}{(2\pi)^3}\right] \;\;
\nonumber \\ && \!\!
+ \int k_3^2k_4^2T_\varphi(\{{\bi k}_j\})
\delta(\sum_{j=1}^4{\bi k}_j)
\prod_{j=1}^4 \frac{d^3{\bi k}_j}{(2\pi)^3},
\nonumber \\ &&
\eeqa
where $T_\varphi$ is the connected trispectrum of the potential in 
the ${\bf x}$ coordinate system.  BMR argues 
that since inflation produces Gaussian initial conditions, the trispectrum 
term drops out, and if $n_s\le 1$, where $P_\varphi(k)\propto k^{n_s-4}$, 
there is an infrared divergence.  We show in the next section that this 
divergence disappears in the context of single-field inflation if one 
calculates the variance of $q_3$ instead of just a single term.

\subsection{Infrared divergences in single-field inflation}
\label{ss:ircancel}

The computation of Var$(\varphi\nabla_{\bf x}^2\varphi)$ is an important 
step in determining the observable deceleration parameter.  Nevertheless, 
we cannot observe $\varphi\nabla_{\bf x}^2\varphi$ directly, rather what 
we observe is $q_3$.  Therefore we want to know Var$\,q_3$.  This section 
examines the contribution of the very long-wavelength modes to Var$\, 
q_3$ predicted by single-field inflation.  We are especially interested in 
the case of $n_s\le 1$, where the variance $I_{-1}\equiv$Var$\,\varphi$ is 
infrared-divergent: $I_{-1}\rightarrow +\infty$.

In this 
section, we retain the terms from Eq.~(\ref{eq:q0y}) that have no factors 
of $k_L$, i.e. we drop $\varphi_{L,i}\varphi_{S,i}$, and are no higher 
than second order in $\varphi_L$, which is the order at which BMR found 
an infrared divergence.  BMR argued that the $\varphi_{L,i}\varphi_{S,i}$
term is subdominant and in any case the variance of $\varphi_{L,i}$ has no 
infrared divergence for $n_s\le -1$.  The equation for $q_3$ is then
\beq
q_3 = \frac{1}{2} + H_0^{-2}\left( \frac{5}{9}
+ \frac{50}{27}\varphi_L + \frac{250}{81}\varphi_L^2 \right)
\nabla_{\bf x}^2\varphi_S.
\label{eq:q3-approx}
\eeq
In order to determine the mean and variance of $q_3$ we need a measure.  
The two most obvious measures are the coordinate measure $d^3{\bi x}$ and 
the measure $dN$ associated with the number $N$ of CDM particles.  Since 
at early times the CDM density is constant in the synchronous comoving 
gauge, it follows that $dN$ is proportional to the volume element at early 
times,
\beq
dN \propto e^{-3\Psi(x^i,t=0)} d^3{\bi x} = e^{-5\varphi(x^i)}d^3{\bi x}.
\eeq
We can unify these two measures by working with the measure $d\mu$ defined 
by
\beq
d\mu = \left[ 1 + \sum_{j=1}^\infty \nu_j
\varphi^j({\bi x}) \right] d^3{\bi x}
\equiv J\, d^3{\bi x},
\eeq
where $\nu_j$ are constants.  The choice $\nu_j=0$ 
corresponds to the coordinate measure, and $\nu_j=(-5)^j/j!$ corresponds 
to the CDM particle number measure, up to an overall constant that does 
not affect the averages.  The function $J$ is the Jacobian.

In order to compute the mean and variance of $q_3$ in 
Eq.~(\ref{eq:q3-approx}), we must compute some averaged values involving 
$\varphi_L$ and $\nabla^2\varphi_S$.  The values linear in $\varphi_S$ 
that we need are
\beq
\langle \nabla_{\bf x}^2\varphi_S\rangle_{\bi x} = 
0,
\eeq
which holds by spatial homogeneity;
\beq
\langle \varphi_L 
\nabla_{\bf x}^2\varphi_S\rangle_{\bi x} = 0,
\eeq
which results from 
$\varphi_L$ and $\varphi_S$ having different wavenumbers; and
\beq
\langle 
\varphi_L^2 \nabla_{\bf x}^2\varphi_S\rangle_{\bi x} = 0,
\eeq
which 
results from the small-scale wavenumbers being very large compared to the 
large-scale wavenumbers so that the triangle inequality cannot be 
satisfied.  Note that these averages are all taken with respect to the 
coordinate measure (we convert to $d\mu$ later).

We will also need the average values $\langle \varphi_L^n (\nabla_{\bi x}
\varphi_S)^2\rangle_{\bi x}$ that are quadratic in $\varphi_S$.  To write 
these, we follow KMNR in defining
\beq
I_\vartheta = \int k^{\vartheta+1} P^{(0)}_{\varphi}(k) \frac{d^3{\bi 
k}}{(2\pi)^3},
\eeq
where $P^{(0)}_{\varphi}(k)$ is the power spectrum of $\varphi$ calculated 
by standard first-order perturbation theory in inflation (i.e. with only 
quadratic terms in the perturbation Lagrangian).  In what follows we will 
consider only the long-wavelength contributions to $I_{-1}$, as these are 
the contributions that produce the infrared divergence reported by BMR 
(and in any case they dominate for $n_s\le 1$).  The first-order theory 
power spectrum is given explicitly by (see e.g. Eq.~(2.20) of
Ref.~\cite{2003JHEP...05..013M})
\beq
P^{(0)}_\varphi(k) = \frac{288\pi^4GH_\ast^4}{25k^3\dot\phi_\ast^2},
\eeq
where $H_\ast$ is the Hubble rate when the mode $k$ becomes super-Hubble 
and $\dot\phi_\ast$ is the rate of change of the inflaton field $\phi$ at 
that time.  (Note that our $\varphi$ is denoted
by $-\frac{3}{5}\zeta$ in Ref.~\cite{2003JHEP...05..013M}.)

The easiest average value to compute is
\beq
\langle \varphi_L^2(\nabla_{\bi x} 
\varphi_S)^2\rangle_{\bi x} = \langle \varphi_L^2 \rangle_{\bi x}\langle 
(\nabla_{\bi x} \varphi_S)^2\rangle_{\bi x} = I_{-1}I_3,
\eeq
which results from decoupling of the long- and short-wavelength modes.  
The integral $I_3$ is dominated by short wavelengths, and (for $n_s\le 1$) 
$I_{-1}$ by long wavelengths.
The average values with 0 or 1 power of $\varphi_L$ are however more 
complicated because of corrections for the non-Gaussianity of $\varphi$.  
The average value with 1 power of $\varphi_L$ and 2 of $\varphi_S$ is 
determined from the bispectrum configuration of $\varphi$ where the 
Fourier-space triangle has one short and two long sides, $k_3\ll 
k_1\approx k_2$.
This configuration has been computed by Ref.~\cite{2003JHEP...05..013M} 
(see Eq.~4.7 in that paper) and is
\beq
\langle\varphi_{{\bi k}_1} \varphi_{{\bi k}_2} \varphi_{{\bi k}_3}\rangle
= \frac{5}{3}[n^{(0)}_{s}(k_1)-1]P_\varphi^{(0)}(k_1) P_\varphi^{(0)}(k_3)
\delta(\sum_{j=1}^3{\bi k}_j),
\label{eq:3pt}
\eeq
where $n^{(0)}_{s}(k_1)$ is the scalar spectral index.
\footnote{Note that ``$n_s$'' in Ref.~\cite{2003JHEP...05..013M} denotes 
what most cosmologists call $n_s-1$.}  This implies
\beq
\langle \varphi_L(\nabla_{\bi x}\varphi_S)^2\rangle_{\bi x} =
\frac{5}{3}I_{-1}\int [n_s^{(0)}(k_1)-1]
k^4 P_\varphi^{(0)}(k)\frac{d^3{\bi k}}{(2\pi)^3}.
\eeq

The average $\langle (\nabla_{\bi x}\varphi_S)^2\rangle_{\bi x}$ at first 
glance appears to be just $I_3$.  However we require the lowest-order 
correction due to long-wavelength perturbations.  Thus one must return to 
the derivation of Eq.~(\ref{eq:3pt}) in order to obtain the correct 
result.  The argument provided by Ref.~\cite{2003JHEP...05..013M} for this 
situation is that the small-scale modes $\varphi_S$ are unaffected by 
large-scale modes except that these large-scale modes re-scale when $k_S$ 
becomes super-Hubble.  In particular, for fixed large-scale perturbation 
$\varphi_L$, the small-scale modes exit the horizon 
$\zeta_L=-\frac{5}{3}\varphi_L$ $e$-folds of inflation earlier than in the 
unperturbed case.  Therefore their power spectrum, conditioned on 
$\varphi_L$, is
\beqa
\frac{k^3}{2\pi^2}P_\varphi(k)|_{\varphi_L} &=& 
\frac{(ke^{-\zeta_L})^3}{2\pi^2}
P^{(0)}_\varphi(ke^{-\zeta_L})
\nonumber \\
&=& 
\frac{(ke^{5\varphi_L/3})^3}{2\pi^2}P^{(0)}_\varphi(ke^{5\varphi_L/3}).
\label{eq:conditional}
\eeqa
If we want the power spectrum of the small-scale modes averaged over all 
regions of space, then one needs to average Eq.~(\ref{eq:conditional}) 
over the probability distribution of $\varphi_L$.  Taylor-expanding 
Eq.~(\ref{eq:conditional}) to second order in $\varphi_L$ and doing the 
averaging yields
\beqa
\frac{k^3}{2\pi^2}P_\varphi(k) &=&
\frac{k^3}{2\pi^2}P^{(0)}_\varphi(k)
\nonumber \\ &&
+ \frac{\langle \varphi_L^2\rangle_{\bi x}}{2}
\frac{d^2}{d\varphi_L^2}\left[
\frac{(ke^{5\varphi_L/3})^3}{2\pi^2}P^{(0)}_\varphi(ke^{5\varphi_L/3})
\right],
\nonumber \\ &&
\label{eq:temppl2}
\eeqa
where the second derivative term comes from the variance of $\varphi_L$, 
and there is no first derivative term because 
$\langle\varphi_L\rangle_{\bi x}=0$.  The variance of $\varphi_L$ is 
$\langle \varphi_L^2\rangle_{\bi x}=I_{-1}$.  If we define the variable
\beqa
\tilde\alpha_s^{(0)}(k) &=& \frac{1}{k^3P^{(0)}_\varphi(k)}
\frac{d^2}{(d\ln k)^2}[k^3P^{(0)}_\varphi(k)]
\nonumber \\
&=& \alpha_s^{(0)}(k) + [n_s^{(0)}(k)-1]^2,
\eeqa
which is related to (but different from) the scalar running 
$\alpha_s^{(0)}(k)$, Eq.~(\ref{eq:temppl2}) simplifies to
\beq
\langle (\nabla^2\varphi_S)^2\rangle =
\int k^4 P^{(0)}_\varphi(k) \left[ 1 + \frac{25}{18}
\tilde\alpha_s^{(0)}(k) I_{-1} 
\right] \frac{d^3{\bi k}}{(2\pi)^3}.
\label{eq:smear}
\eeq
Physically, Eq.~(\ref{eq:smear}) represents the ``smearing'' of the 
relation between $k$ and physical scale due to the variance of 
$\varphi_L$.  This smears the power spectrum $P_\varphi(k)$.  Thus it is 
not surprising that Eq.~(\ref{eq:smear}) contains a second derivative of 
the power spectrum.  The smearing effect can be thought of as a loop 
correction to the scalar power spectrum in the sense that it involves an 
integration over an undetermined momentum (that of $\varphi_L$, which 
is packaged into $I_{-1}$ here).

The above results are most easily expressed if we define the integrals
\beq
I_{\vartheta,\beta} = \int k^{\vartheta-2}
\left\{ \partial_{\ln k}^\beta [k^3P_\varphi^{(0)}(k)] \right\}
\frac{d^3{\bi k}}{(2\pi)^3}.
\label{eq:ivb}
\eeq
In particular,
\beq
I_{3,1} = \int k^4 [n_s^{(0)}(k)-1] P^{(0)}_\varphi(k)
\frac{d^3{\bi k}}{(2\pi)^3}
\eeq
and
\beq
I_{3,2} = \int k^4 \tilde\alpha_s^{(0)}(k) P^{(0)}_\varphi(k)
\frac{d^3{\bi k}}{(2\pi)^3}.
\eeq
Then the average values containing $(\nabla_{\bi x}^2\varphi_S)^2$ are
\beqa
\langle (\nabla^2\varphi_S)^2\rangle &=&
I_3 + \frac{25}{18}I_{3,2}I_{-1},
\nonumber \\
\langle\varphi_L(\nabla^2\varphi_S)^2\rangle &=&
\frac{5}{3}I_{3,1}I_{-1}, {\rm ~~and}
\nonumber \\
\langle\varphi_L^2(\nabla^2\varphi_S)^2\rangle &=& I_3I_{-1}.
\eeqa

Finally we come to the issue of the mean and variance of $q_3$.  The mean 
of any quantity $X$ with respect to the measure $\mu$ is related to the 
Jacobian $J$ via
\beq
\langle X\rangle_\mu = \frac{\langle XJ\rangle_{\bi x}}{\langle 
J\rangle_{\bi x}}.
\eeq
We have evaluated $q_3$ including all terms of order 
$\varphi_L^a\varphi_S^b$, where $a\le 2$ and $b\le 1$.  Therfore we can 
calculate the mean $\langle q_3\rangle$ only up to order 
$\varphi_L^2\varphi_S$.  In principle with 
the Taylor series cut off in this way we can only evaluate Var$\,
q_3$ to this order as well.  However, if one switches the variables 
to $Q_3=q_3-\frac{1}{2}$, then it turns out that the Taylor expansion of 
$Q_3$ in $\varphi_L$ and $\varphi_S$ contains no terms zeroeth order in 
$\varphi_S$.  This fact allows us to compute Var$\,
Q_3$ and hence Var$\, q_3$ to order $\varphi_L^2\varphi_S^2$.

We now calculate $\langle q_3\rangle_\mu$ to first order in $\varphi_S$ 
and $\varphi_L$:
\beq
\langle q_3\rangle_\mu = \frac{1}{2} + H_0^{-2}\frac{
\langle J(
\frac{5}{9} + \frac{50}{27}\varphi_L + \frac{250}{81}\varphi_L^2
)\nabla_{\bi x}^2\varphi_S
 \rangle_{\bi x}
}{\langle J\rangle_{\bi X}}.
\label{eq:temp1}
\eeq
Now if we are dropping all terms second order in $\varphi_S$, then $J$ 
can be considered a function of $\varphi_L$ for the purposes of 
evaluating the numerator.  Thus the numerator in Eq.~(\ref{eq:temp1}) 
is the mean value of $\nabla_{\bi x}^2\varphi_S$ times a function 
of $\varphi_L$.  Since we are working only to order $\varphi_L^2$, and 
we know that $\langle\varphi_L^n\nabla_{\bi x}^2\varphi_S\rangle_{\bi 
x}=0$, it follows that the numerator in Eq.~(\ref{eq:temp1}) vanishes.  
Therefore
\beq
\langle q_3\rangle_\mu = \frac{1}{2} + O(k_L,\varphi_S^2,\varphi_L^3).
\label{eq:mean-q3}
\eeq

We will compute Var$\,q_3$ to first order in the power 
spectrum of the long-wavelength modes or to second order in the 
long-wavelength modes themselves, i.e. to order $\varphi_L^2$.  We will 
also work to second order in $\varphi_S$.  Using Eq.~(\ref{eq:mean-q3}), 
and defining $Q_3=q_3-\frac{1}{2}$, we find
\beq
{\rm Var}\,q_3 = \langle Q_3^2\rangle_\mu - \langle Q_3\rangle_\mu^2
= \langle Q_3^2\rangle_\mu - O(k_L^2,\varphi_S^4,\varphi_L^6).
\eeq
That is, the term $\langle Q_3\rangle_\mu^2$ cannot contribute at order
$\varphi_S^2$
even though we have only computed $\langle Q_3\rangle_\mu$ to order 
$\varphi_S$.  Then
\beq
{\rm Var}\,q_3 = \frac{\langle JQ_3^2\rangle_{\bi x}}
{\langle J\rangle_{\bi x}}.
\label{eq:varq3}
\eeq
The denominator is
\beq
\langle J\rangle_{\bi x}
= \langle 1 + \nu_1(\varphi_L+\varphi_S) + 
\nu_2(\varphi_L^2+2\varphi_L\varphi_S) \rangle_{\bi x}
= 1 + \nu_2 I_{-1}.
\label{eq:q3den}
\eeq
The numerator is
\beqa
\langle JQ_3^2\rangle_{\bi x} &=&
\frac{25I_3}{81H_0^4} + \frac{625I_{3,2}I_{-1}}{1458H_0^4}
+ \frac{2500I_{3,1}I_{-1}}{729H_0^4}
\nonumber \\ &&
+ \frac{5000I_3I_{-1}}{729H_0^4}
+ \frac{125\nu_1I_{3,1}I_{-1}}{243H_0^4}
\nonumber \\ &&
+ \frac{500\nu_1I_3I_{-1}}{243H_0^4}
+ \frac{25\nu_2I_3I_{-1}}{81H_0^4}.
\label{eq:q3num}
\eeqa
The physical origin of the terms in Eq.~(\ref{eq:q3num}) is as follows: 
the first term is the usual variance of the deceleration due to some 
patches of the universe being over- or under-dense.  The second term is 
the ``smearing'' loop correction described above.  The third term is a 
correlation between the order $\nabla_{\bi x}^2\varphi_S$ and 
$\varphi_L\nabla_{\bi x}^2\varphi_S$ in $q_3$ (Eq.~\ref{eq:q3-approx}) 
that arises from the nonvanishing bispectrum from inflation.  The fourth 
term comes from two places: the variance of the $\varphi_L\nabla_{\bi 
x}^2\varphi_S$ term (which is the infrared-divergent term identified by 
BMR) and the correlation between $\nabla_{\bi x}^2\varphi_S$ and 
$\varphi_L^2\nabla_{\bi x}^2\varphi_S$ terms in Eq.~(\ref{eq:q3-approx}).  
In this case, the variance term contributes a coefficient of 
$\frac{2500}{729}$ and the correlation contributes $\frac{2500}{729}$, 
yielding the total coefficient of $\frac{5000}{729}$.  The fifth, sixth, 
and seventh terms represent the modulation of the earlier terms by the 
non-coordinate measure.

Looking at Eqs.~(\ref{eq:q3den}) and (\ref{eq:q3num}), it appears at first 
glance that $I_{-1}$ affects Var$\, q_3$ in a highly nontrivial way 
depending on the values of $I_3$, $I_{3,1}$, and $I_{3,2}$.  This is in 
fact not the case, because $I_3$, $I_{3,1}$, and $I_{3,2}$ are not 
independent quantities.  Integration by parts in the definition 
(Eq.~\ref{eq:ivb}) and dropping boundary terms gives
\beq
I_{\vartheta,\beta} = -(\vartheta+1)I_{\vartheta,\beta-1}.
\label{eq:recursion}
\eeq
Noting that the definitions imply $I_3=I_{3,0}$, we conclude that 
$I_{3,1}=-4I_3$ and $I_{3,2}=16I_3$.  One might object that the boundary 
terms cannot be neglected because $I_3$ is ultraviolet divergent.  Of 
course, if this divergence is not regulated, $I_3=\infty$, in which case 
even the first-order perturbation theory result for Var$\, q_3$ is 
infinite and it makes little sense to talk about higher-order corrections.  
Physically, the divergence is regulated by putting in some cutoff in 
wavenumber $k$.  In the case of KMNR, the regulator is placed at the 
horizon scale, $k\sim H_0$, where one separates super-Hubble 
perturbations from the sub-Hubble modes (the latter are considered to be 
observable perturbations to the universe, rather than a correction to 
the observed scale factor).  In the case of deceleration parameter 
measurements from SNe Ia, the cutoff scale is roughly of the order of the 
distance $D$ to the supernovae; modes with wavelengths less than this 
cannot be treated by a local deceleration parameter based on the second 
derivative of the $d_L$-$z$ relation at $z=0$.  In either case, the cutoff 
occurs at a fixed physical scale (the horizon scale or the distance to the 
supernovae), rather than at a fixed coordinate scale ${\bi x}$.  Therefore 
the position of the cutoff in terms of coordinate wavenumber $k$ is 
modulated by $\varphi_L$ as $k_{\rm cutoff}\propto e^{-5\varphi_L/3}$.  
Since this is the same as the modulation of the wavenumber in the 
inflationary power spectrum $P_\varphi(k)$ by large-scale modes (see 
Eq.~\ref{eq:conditional}), one can simply absorb the cutoff into 
$P^{(0)}_\varphi(k)$.  Then $I_3$ becomes finite and 
Eq.~(\ref{eq:recursion}) is valid.

Substituting our results that $I_{3,1}=-4I_3$ and $I_{3,2}=16I_3$ into 
Eq.~(\ref{eq:q3num}) reduces it to
\beq
\langle JQ_3^2\rangle_{\bi x} = \frac{25I_3}{81H_0^4}(1+\nu_2I_{-1})
\eeq
and hence
\beq
{\rm Var}\, q_3 = \frac{25I_3}{81H_0^4},
\eeq
independently of the value of $I_{-1}$.  The long-wavelength contribution 
to $I_{-1}$, which is responsible for the infrared-divergent terms, 
cancels out when all contributions are included.  This is true for both 
the coordinate and the particle number measure.
This specific example 
confirms the general conclusions based on causality found above.  

\subsection{Interpretation}
\label{ss:3117}

The correspondance between our results and BMR reveals the origin of the 
infrared divergence: it comes from the assumption that $\varphi_S$ was 
taken as a function of ${\bf x}$ rather than ${\bf y}$.  Since the ${\bf 
x}$ coordinate system is distorted by superhorizon perturbations, these 
superhorizon modes distort structures within the horizon and affect the 
luminosity distance.  Whether this effect is observable depends on whether 
the values of $\varphi_S$ within the horizon are a Gaussian random field 
with the usual power spectrum $P_\varphi(k)$ in ${\bf x}$ or in ${\bf y}$.  
This can be answered only by a theory of the initial conditions. If the 
initial conditions are set by single-field inflation, then the 
fluctuations $\varphi_L$ that determine the relation between the 
coordinates ${\bf x}$ and ${\bf y}$ are set down when these large scales 
leave the horizon.  They then become classical, and later on (i.e. many 
$e$-folds of inflation later) the perturbations $\varphi_S$ are generated.  
Since the generation of $\varphi_S$ must be causal, one would expect that 
within regions small compared to the wavelength of $\varphi_L$, inflation 
generates $\varphi_S$ homogeneous and isotropic in the ${\bf y}$ 
coordinate system.  If the calculation is done in the ${\bf x}$ coordinate 
system, as we did in Sec.~\ref{ss:ircancel}, then the infrared divergence 
from the second-order perturbation theory found by BMR cancels against 
three other infrared divergences: one arising from the correlation of 
first- and third-order perturbation terms, one from the correlation of 
first- and second-order perturbation terms that arises from the primordial 
bispectrum, and one from loop corrections to the power spectrum predicted 
by inflation.  Physically, the disappearance of $I_{-1}$ from the 
statistical properties of observables such as Var$\, q_3$ is a 
manifestation of the fact that inflation wipes out initial conditions: the 
later stages of inflation prevent one from observing the pre-existing 
larger-scale structure of the universe, including the perturbations 
generated during the early stages of inflation.  (Recall that this is also 
how inflation solves the flatness and homogeneity problems.)

The scenario proposed by KMNR is closely related to the infrared 
divergence. Specifically, they argue that the 
$e^{10\varphi_L/3}\nabla^2_{\bf x}\varphi_S$ term in Eq.~(\ref{eq:q0y}) 
has a large variance that causes the deceleration parameter to also have a 
large variance.  At second order in perturbation theory, this term is the 
infrared-divergent $\varphi_L\nabla^2_{\bf x}\varphi_S$ found by BMR, 
which we find to be canceled by other divergences that were not considered 
by BMR.  KMNR uses the full prefactor $e^{10\varphi_L/3}$ and hence 
includes the divergence associated with the $\varphi_L^2\nabla^2_{\bf 
x}\varphi_S$ term, but does not consider the divergences from the 
bispectrum or the loop correction to the power spectrum.  From the 
perturbative calculation of Sec.~\ref{ss:ircancel}, it is not obvious 
whether this cancellation extends to arbitrary order.  However if the 
later $e$-folds of inflation produce perturbations homogeneous and 
isotropic with power spectrum $P^{(0)}_\varphi(k)$ in the ${\bf y}$ 
coordinate system, as must happen on account of causality, then we need 
only consider Eq.~(\ref{eq:q0}) to realize that the superhorizon structure 
is irrelevant, and the cancellation of the infrared divergences that arise 
in the ${\bf x}$ coordinate system must be exact.

[As noted in Sec.~\ref{sec:decel}, KMNR use $q_1$ rather than $q_3$ as 
their deceleration parameter.  As far as Eq.~(\ref{eq:q0y}) is concerned, 
this does not matter since Eq.~(\ref{eq:q3-1st}) is valid for $q_1$ as 
well as for $q_3$.  To see this, note that in first order perturbation 
theory, if one does a spherical expansion of the perturbation around the 
observer, symmetry implies that only the $l=0$ modes contribute to $q_1$ 
or $q_3$.  If only the $l=0$ modes are present then we have $q_{3'}$ 
independent of direction $n^i$ and $q_3=q_4$.  Since $H_1=H_4$, a 
comparison of Eqs.~(\ref{eq:hq1}) and (\ref{eq:hq2}) with the assumption 
that $\sigma=\omega=0$ from spherical symmetry implies $q_1=q_4$.  We then 
have $q_1=q_3$ by transitivity.  Hence Eq.~(\ref{eq:q3-1st}) and thus 
Eq.~(\ref{eq:q0y}) are valid for $q_1$ instead of $q_3$.]

\section{Conclusions}

In this paper, we have investigated the KMNR explanation for the 
accelerating expansion of the universe, which suggests that (i) in 
sufficiently underdense regions of the universe, the Hubble expansion 
appears to accelerate ($q_1<0$) even with only normal matter present and 
Einstein gravity; and (ii) the variance Var$\,q_1$ of the deceleration 
parameter is much greater than the simple calculation $\sim (10^{-5})^2$ 
because of the influence of perturbation modes with wavelengths many 
orders of magnitude larger than the Hubble length, so that the 
acceleration $q_1<0$ has a non-negligible probability of actually 
occurring.  We have shown that suggestion (i) is ruled out by the 
Raychaudhuri equation, i.e. with GR and the SEC one always has $q_1\ge 0$.  
We have also shown that suggestion (ii) is not true for the perturbations 
generated by single-field inflation, which does indeed predict Var$\,q_1$ 
of order $I_3^2\sim (10^{-5})^2$ (assuming an ultraviolet cutoff at the 
Hubble scale, as in KMNR).  Therefore the KMNR mechanism cannot explain 
the acclerating expansion of the universe.

What freedom is there to construct models related to KMNR that do explain 
the acceleration of the universe with superhorizon perturbations?  Of the 
two elements of the KMNR mechanism, (ii) may be the easiest to circumvent.  
In single-field inflation, the perturbations laid down by the early stages 
of inflation are adiabatic so the perturbations laid down by the later 
stages of inflation must ``look the same'' (i.e. have the same power 
spectrum in our ${\bi y}$ coordinate system) everywhere.  This requirement 
is the physics underlying the particular limiting forms for the bispectrum 
and the one-loop correction to the power spectrum that we used in 
Sec.~\ref{ss:ircancel} to argue that single-field inflation produces no 
infrared divergences.  In contrast, multi-field inflation models can 
produce isocurvature modes so that different patches of the universe look 
different.  In this case Var$\, q_1$ may be large because one takes the 
variance over regions with a different mix of cosmological fluids 
\cite{2004astro.ph.10541K}.  Even then, however, an observer can only see 
one Hubble volume because of causality and so Var$\, q_1$ is not an 
observable, rather it is the variance of a distribution from which one 
obtains a single sample.  In such a case the only role played by the 
superhorizon perturbations is to alter the initial conditions: all 
observables, including $q_1$, are obtained from causal evolution that can 
be calculated from knowledge of the perturbations within the observer's 
horizon.  More importantly for the dark energy question, the combination 
of GR and the SEC still forbids $q_1<0$.

If one is to find a way to keep GR, the SEC, zero vorticity, and neglect 
perturbations with wavelengths small compared to the Hubble length, and 
still maintain consistency with the observational results, one is forced 
to find a situation in which $q_1$ as defined above is not what is 
actually measured in the SN Ia experiments.  BMR argued that $q_3$ is a 
better representation of what is observed, since it is based on luminosity 
distances; the concordance cosmology has $q_3\sim -0.6$.  Unfortunately, 
we found in Sec.~\ref{sec:decel} that $q_3<0$ is possible within the 
GR+SEC framework only for anisotropic expansion, since the angular average 
$\langle H_{3'}^2q_{3'}\rangle_{4\pi}$ must be non-negative.  In this 
case, there must also be lines of sight along which $q_{3'}>0$.  There is 
no observational evidence that this is the case. Indeed, it would be an 
extraordinary mystery to have the deceleration parameter $q_{3'}$ vary by 
an amount of order unity on scales comparable to the Hubble length and 
still produce a CMB isotropic to the level of a few parts in $10^5$.

We conclude that cosmological models based on GR with irrotational initial 
conditions and perturbations only at and above the Hubble scale and only 
matter fields that conform to the SEC cannot explain the accelerating 
expansion.  This paper does not exclude the possibility of using 
backreaction from sub-Hubble perturbations to explain the accelerating 
expansion.  The latter possibility is difficult to investigate as it 
involves complicated nonlinear physics; the perturbative calculations 
\cite{1996ASPC...88..267S}, which account for the nonlinear evolution of 
density perturbations, but not for strong field GR effects, suggest that 
the sub-Hubble backreaction is small \footnote{The possibly 
ultraviolet-divergent term in Ref.~\cite{2004JCAP..402..003R} was found by 
Ref.~\cite{2005PhRvD..71b3524K} to not significantly affect the expansion 
rate, at least at second order.}. Nevertheless, only a full 
non-perturbative analysis would be definitive since there are rare regions 
of the universe such as black holes that cannot be described as a 
perturbation of an FRW spacetime.  Regardless of the sub-Hubble physics, 
however, the superhorizon perturbations are not a viable mechanism to 
explain the acceleration of the universe: evolution of perturbations that 
lie beyond the horizon in real space cannot affect observables, and the 
superhorizon perturbations can only act via their effects on the initial 
conditions within our horizon.  These effects are then constrained by the 
``no go'' theorems that require $q_1$ and $q_4$ to be non-negative unless 
one invokes vorticity, modified gravity, or dark energy.

\begin{acknowledgments}

C.H. wishes to thank Latham Boyle and Mustapha Ishak for enlightening 
discussions.  C.H. is supported by NASA NGT5-50383.
 U.S. is supported by Packard Foundation, NASA NAG5-11489
and NSF CAREER-0132953.

\end{acknowledgments}

\appendix

\section{Convergence radius of series for $\Psi$}
\label{app:conv}

The purpose of this appendix is to investigate the convergence properties 
of the power series
Eq.~(\ref{eq:psi-open}) for $\Psi$ in the 
model of Sec.~\ref{sec:accel}.  Our framework will be the result from 
complex analysis that a Taylor series of an analytic function has a radius 
of convergence given by 
the distance to the nearest singularity.  We begin by defining the 
dimensionless scale factor
\beq
\beta = \frac{Ca}{H_0^2} = \left(\frac{3t}{4\Xi}\right)^{2/3},
\eeq
so that Eq.~(\ref{eq:eta}) becomes
\beq
a_{\rm open} = \Xi C^{1/2}(\cosh\eta-1)
\eeq
and
\beq
\beta^{3/2} = \frac{3}{4}(\sinh\eta-\eta).
\label{eq:beta32}
\eeq
It will be most convenient for the purposes of this appendix to treat
Eq.~(\ref{eq:psi-open}) as a Taylor series in the dimensionless 
variable $\beta$ rather than $a$ or $C$.  If we determine the radius of 
convergence in $\beta$, then this immediately yields the radius of 
convergence in $a$ or $C$.

We want to determine the singularities of the function $\Psi$ of 
Eq.~(\ref{eq:po}) or equivalently $\ln (a_{\rm open}/a)$.  First we 
comment on the analytical structure near $\beta=0$.  The function
$\eta(\beta)$ defined by Eq.~(\ref{eq:beta32}) is double-valued near zero, 
with two solutions $\eta_1$ and $-\eta_1$.  This presents no problem 
because both result in a single value for $a_{\rm open}$.  Specifically,
near $\beta=\eta=0$, we have $\frac{3}{4}(\sinh\eta-\eta)\sim 
\frac{1}{8}\eta^3$ so 
$\eta\sim 2\beta^{1/2}$ and
\beq
a_{\rm open}\sim \Xi C^{1/2}\frac{\eta^2}{2}\sim 2 \Xi C^{1/2} \beta
\sim \frac{H_0^2}{C}\beta.
\label{eq:aob}
\eeq
Thus $a_{\rm open}$ is well-behaved at the origin (even for complex 
values).
  
Equation~(\ref{eq:psi-open}) is the Taylor expansion of Eq.~(\ref{eq:po}) 
in $\beta$ and hence the value of $\beta$ with smallest absolute value 
that makes $\ln (a_{\rm open}/a)$ singular determines the radius of 
convergence of Eq.~(\ref{eq:psi-open}).  We can find these singularities 
by taking the derivative,
\beqa
\frac{d}{d\beta}\ln\frac{a_{\rm open}(\beta)}{a(\beta)}
&=& \frac{d}{d\beta}\ln\frac{\cosh\eta-1}{\beta}
\nonumber \\
&=& \frac{\sinh\eta}{\cosh\eta-1}\frac{d\eta}{d\beta} - \beta^{-1}.
\eeqa
Using the implicit derivative of Eq.~(\ref{eq:beta32}),
\beq
\frac{3}{2}\beta^{1/2} = \frac{3}{4}(\cosh\eta-1)\frac{d\eta}{d\beta},
\eeq
we find
\beq
\frac{d}{d\beta}\ln \frac{a_{\rm open}}{a}
= \frac{2\sinh\eta}{(\cosh\eta-1)^2}\beta^{1/2}-\beta^{-1}
\label{eq:ddb}
\eeq
This function appears to have singularities where $\cosh\eta=1$ or 
$\beta=0$.  The singularity at $\beta=0$ is only apparent since at small 
values of $\beta$, we know that $\ln(a_{\rm open}/a)\rightarrow 0$.  Thus 
the only singularities can appear when $\cosh\eta=1$ and $\beta\neq 0$.  
The solutions to $\cosh\eta=1$ are $\eta=-2\pi i m$ where $m$ is any 
integer.  Any point with $\eta=-2\pi i m$, $m\neq 0$ must actually be a 
singularity since for $\eta=-2\pi i m+\epsilon$, we have (by periodicity 
of the hyperbolic functions with period $2\pi i$)
\beq
\frac{2\sinh\eta}{(\cosh\eta-1)^2}
= \frac{2\sinh\epsilon}{(\cosh\epsilon-1)^2}
\sim \frac{2\epsilon}{(\epsilon^2/2)^2}\sim 8\epsilon^{-3};
\eeq
then since $\beta^{1/2}$ and $\beta^{-1}$ are analytic for $\beta\neq 0$ 
there must be a singularity in Eq.~(\ref{eq:ddb}).  The points with 
$\eta=-2\pi i m$, $m\neq 0$ correspond to
\beq
\beta^{3/2} =\frac{3}{4}[\sinh(-2\pi i m)+2\pi i m]
= \frac{3\pi i m}{2}.
\eeq
The values $m=0$ correspond to $\beta=0$ where there is no singularity.  
The other singularities occur at the points
\beq
\beta = \left(\frac{3}{2}\pi\right)^{2/3} m^{2/3} E,
\eeq
$m\neq 0$, where $E$ is any cube root of $-1$.  The closest such 
singularities to the origin have $m=\pm 1$ and hence 
$|\beta|=(\frac{3}{2}\pi)^{2/3}$.  Thus we conclude that the radius of 
convergence 
of Eq.~(\ref{eq:psi-open}) is $|\beta|<(\frac{3}{2}\pi)^{2/3}$ or
\beq
|CH_0^{-2}a|<\left(\frac{3}{2}\pi\right)^{2/3}.
\eeq

\end{document}